\documentclass[%
 reprint,
 amsmath,amssymb,
 aps,longbibliography
]{revtex4-1}

\usepackage{graphicx}
\usepackage{dcolumn}
\usepackage{bm}

\begin{document}

\title{Superconductor to Insulator Transition Tuned by Random Gauge Fields}

\author{H. Q. Nguyen}
\altaffiliation{Currently at: Nano and Energy Center, Hanoi University of Science, Vietnam National University, Hanoi, Vietnam
}
\affiliation{Department of Physics, Brown University, Providence, RI 02912}
\author{S. M. Hollen}
\altaffiliation{Currently at: Department of Physics, University of New Hampshire, Durham, NH 03824}
\affiliation{Department of Physics, Brown University, Providence, RI 02912}
\author{J. Shainline}
\altaffiliation{Currently at: National Institute of Standards and Technology, 325 Broadway, Boulder, Colorado, 80305, USA}
\affiliation{School of Engineering, Brown University, Providence, RI 02912}
\author{J. M. Xu}
\affiliation{School of Engineering, Brown University, Providence, RI 02912}
\author{J. M. Valles Jr.}
\affiliation{Department of Physics, Brown University, Providence, RI 02912}

\begin{abstract}
Typically the disorder that alters the interference of particle waves to produce Anderson localization is  potential scattering from randomly placed impurities.  Here we show that disorder in the form of  random gauge fields that act directly on particle phases can also drive localization.  We present evidence of a  superfluid bose glass to insulator transition at a critical level of this gauge field disorder in a nano-patterned array of amorphous Bi islands.  This transition shows signs of metallic transport near the critical point characterized by a resistance $\sim \frac{1}{2}\frac{h}{4e^2}$, indicative of a quantum phase transition. The critical disorder also depends on interisland coupling in agreement with recent Quantum Monte Carlo simulations. Finally, these experiments are uniquely connected to theory because they employ a method for controlling a disorder parameter that coincides directly with a term that appears in model Hamiltonians.  This correspondence will enable further high fidelity comparisons between theoretical and experimental studies of disorder effects on quantum critical systems.
\end{abstract}

\maketitle
A random gauge field adds random increments to the phase of a particle as it traverses a system.  It  appears as a random phase factor in the site to site tunneling integral in tight binding models.  For the most familiar gauge field, a random magnetic field with zero mean, the phase shifts take the form $A_{ij}=\frac{2\pi q}{h}\int_i^j\bf A\cdot d\bf l$ for a particle of charge q moving from $i$ to $j$ in a magnetic vector potential, $\bf A$.  The effects of random gauge fields, also called gauge field disorder, have been considered in attempts to describe anomalous transport in the normal state of high temperature superconductors\cite{LeeRMP06}, graphene\cite{MorozovPRL97,CastroNetoRMP09}, the $\nu=1/2$ state in two dimensional electron gases\cite{HalperinPRB93,MancoffPRB95}, and photons in solid state structures\cite{mittalPRL}.  

Fluctuations in gauge fields influence fermions and bosons distinctly.  Magneto-transport experiments on rippled graphene suggest that they counteract Anderson localization of fermions\cite{MorozovPRL97,CastroNetoRMP09,levyScience2010}.  Similarly, models show that random Chern-Simons gauge fields produce the nearly metallic rather than localized transport associated with the $\nu=1/2$ state\cite{KalmeyerPRB92, HalperinPRB93}.   On the other hand, gauge field fluctuations appear to destroy superfluidity and tend to localize bosons\cite{LeeRMP06}.   Efforts to explain the normal state transport of high $T_c$ superconductors using resonating valence bond models have led investigators to consider the effects of random gauge fields on bosons in two dimensions\cite{LeeRMP06}. Fluctuations in the gauge field appear to suppress Bose condensation and thus, superfluidity at finite temperatures in those treatments of t-J models\cite{LeeRMP06}. 

There have been many efforts to manipulate and engineer gauge fields to address new physics\cite{guineaNP2010,MancoffPRB95, XiePRL98, nogaretJP10}.  A few have applied spatially random magnetic fields to two dimensional electron systems\cite{MancoffPRB95, XiePRL98, nogaretJP10} to investigate models of the $\nu=1/2$ fractional quantum hall state.  The motivation to create ever more versatile quantum simulators of many body systems has led to methods for producing artificial gauge fields in uncharged systems, such as cold neutral atom or quantum optics \cite{dalibardRMP11, aidelsburgerPRL11,mittalPRL}.  Particularly germane to the current report, a couple of groups created disordered gauge fields in Josephson Junction Arrays (JJA).  They fabricated arrays with positional disorder to produce a random amount of flux per plaquette in the presence of a transverse field\cite{forresterPRB88, yun2006experimental}.   Their studies focussed on the effects of this disorder on the classical Berezinski-Kosterlitz-Thouless transition\cite{GranatoPRB86,korshunov2006phase}.  Here, we employ a similar approach to investigate the effects of random gauge fields on the quantum superconductor to insulator transition.  We present  evidence that strengthening a random gauge field weakens a superfluid state: the experiments show that a random gauge field can drive a low superfluid density superconductor into an insulating phase. 

\begin{figure*}[htb]
\begin{center}
\includegraphics[width=5 in,keepaspectratio]{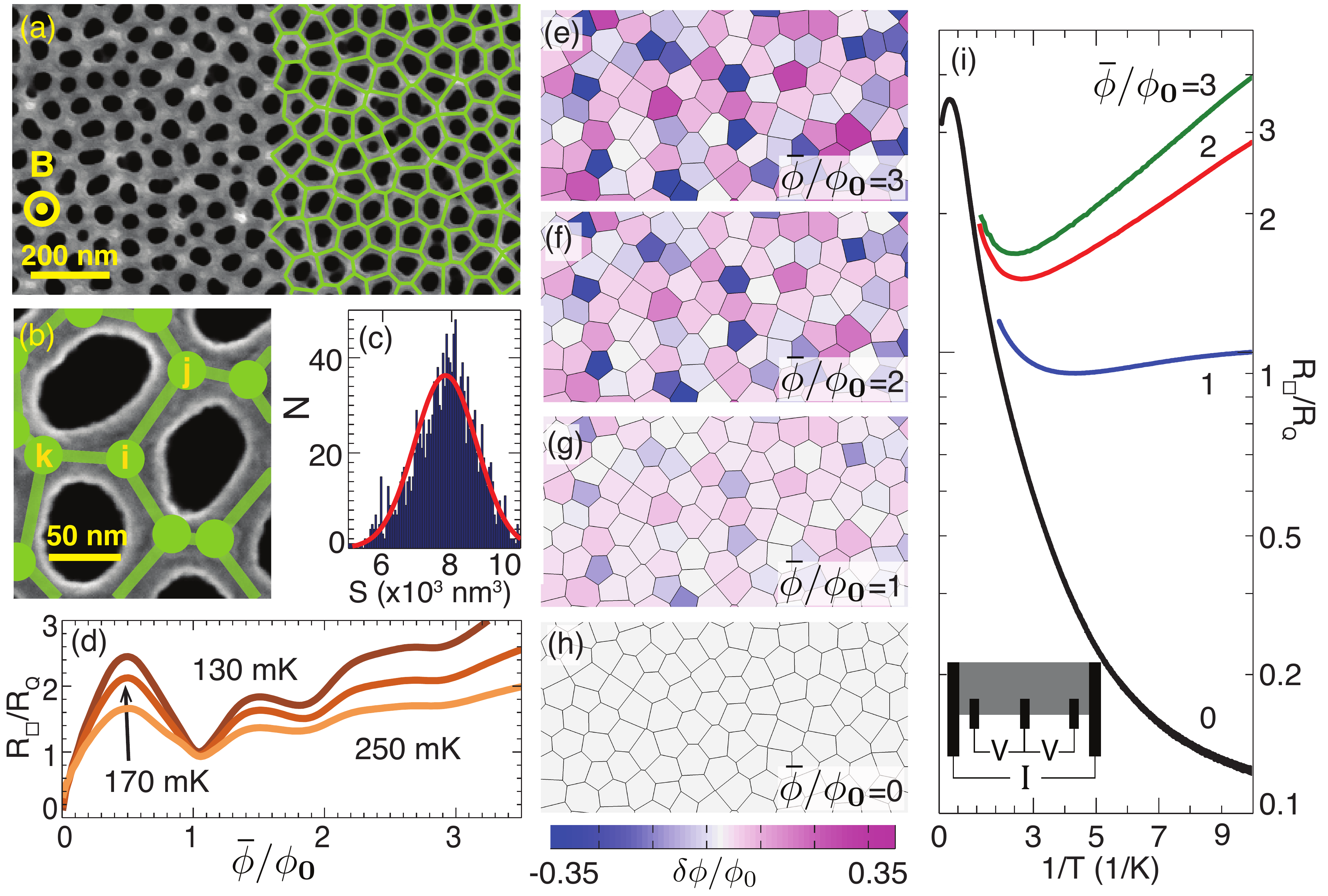}
\caption{Random Gauge Field Tuning Method. a) Scanning electron microscope image of an amorphous Bi nano-honeycomb film. The overlaid green network links defining individual array cells were obtained using a triangulation method.  An applied magnetic field $\mathbf B$ points out of the page. b) Magnified region of a) showing dots to denote nodes. c) Distribution of cell areas defined by the links. d) Low temperature magnetoresistance, normalized to $R_Q$, of a film with normal state sheet resistance of $R_N=20 k\Omega$ plotted versus $\bar\phi/\phi_0$ for three different temperatures. e-h) Maps of the deviation of the magnetic flux through a cell from the average value,  $\delta\phi$, in units of the flux quantum, $\phi_0$, for commensurate fields $\bar\phi/\phi_0=$ 0, 1, 2 and 3.  The random variations in $\delta\phi$ imply random variations in the line integral of the gauge field $A_{ij}$ along links.   This disorder grows proportionally with $\bar\phi/\phi_0$.  i)  Sheet resistance as a function of inverse temperature at commensurate fields for a film with $R_N=20$ k$\Omega$. Inset: Sample measurement setup.}
\label{Fig1}
\end{center}
\end{figure*}

Our investigations employ films patterned into arrays that are on the superconducting side of a thickness tuned superconductor to insulator transition (Fig. 1a)\cite{StewScience07}.   It is helpful to consider their behavior in the light of the quantum rotor model that is commonly used to describe the SIT\cite{dobrosavljevic12,fazioPR01,kimPRB08,swansonPRX14}.  Its Hamiltonian is given by: 
\begin{equation}
H=U\sum_i n_i^2-J\sum_{<ij>}\cos(\theta_i-\theta_j-A_{ij}).
\end{equation}
$n_i$, the number operator for Cooper pairs and $\theta_j$, the phase operator on node $j$ satisfy $[n_i,\theta_j]=i\delta_{ij}$ (see Fig. 1b). The first term is an onsite Coulomb energy of strength $U$ that tends to localize Cooper pairs to individual nodes.   The second term, which sums over nearest neighbors, competes with the first by promoting phase coherence  and a delocalized superfluid state.  The internode coupling $J$ is proportional to the amplitude of the superconducting order parameter on the nodes and tunneling coupling between nodes.  The argument of the cosine is the gauge invariant phase shift, $\eta_{ij}=\theta_i-\theta_j-A_{ij}$, for a boson tunneling directly from island $i$ to island $j$.  In zero magnetic field and for perfectly ordered arrays, this model exhibits a superconductor to insulator transition at a critical coupling $K_c=(J/U)_c=0.206$ \cite{kimPRB08,Nguyen2010} below which quantum phase fluctuations drive Cooper pair localization.  Moreover, this transition occurs at the same critical coupling at all commensurate magnetic field values for which $\sum A_{ij}=2\pi n$ around a plaquette or the number of flux quanta per plaquette $\phi/\phi_0$ is an integer, $n$\cite{fazioPR01,kimPRB08}.   $\phi_0=h/2e$ is the superconducting flux quantum.

Calculations show that adding geometrical disorder to an array in a magnetic field induces a random gauge field that modifies the forgoing ordered array behavior \cite{kimPRB08}. To see how this occurs, consider the amplitude for a Cooper pair tunneling from site $i$ to $j$.  The tunneling probability amplitude from $i$ to $j$ is given by the superposition of all paths connecting them.  These paths interfere constructively to give the greatest net amplitude when $\eta_{ij}=2\pi n$, for integer $n$, along every link in the array.  This condition holds for ideal ordered arrays at commensurate fields. In arrays with a distribution of unit cell areas (like Fig. 1c), however, it is only possible to approximate commensurability by making $\bar\phi/\phi_0=n$ for the average flux per plaquette.  At this average condition, the $A_{ij}$ vary randomly in space to produce gauge invariant phase shifts that deviate randomly from multiples of $2\pi$. The random phase shifts reduce the constructive interference effects described above to make the system more susceptible to phase fluctuations.  Consequently, increases in gauge field disorder are expected to lead to an increase in the critical coupling $K_c$ in order to counteract the enhanced phase fluctuations.  Here we show not only that $K_c$ increases with also that gauge field disorder can be sufficient to drive a superconductor to insulator transition.

We engineered a random gauge field by patterning films into a geometrically disordered hexagonal array (Fig. 1a) and subjecting them to a perpendicular magnetic field.  We thermally evaporated Sb and then Bi onto a cryogenically cooled anodized aluminum oxide substrate with a surface perforated by a disordered triangular array of holes\cite{NguyenPRBRC15}. Similarly produced nano-honeycomb (NHC) films undergo a localized Cooper pair to superfluid transition with increasing deposition \cite{StewScience07}. The nodes (Fig. 1b), which have a relatively larger thickness than the links due to undulations in the substrate surface, harbor more Cooper pairs compared to the links connecting them\cite{hollenPRB11}. Sheet resistance, $R_\square$, measurements were made on a 1 $mm^2$ area of film using standard four point low frequency techniques (inset Fig. 1)\cite{StewScience07}.  The low temperature magneto-resistance of these films  showed a decaying oscillation pattern with a period of $\bar\phi/\phi_0=1$ (Fig. 1d)\cite{NguyenPRBRC15}.  

We characterize the strength of the gauge field disorder by the variance in the distribution of $A_{ij}$, $\Delta A_{ij}$, and an associated phase randomization length, $L_\theta$.   $\Delta A_{ij}$ can be related to the variance in the flux per unit cell\cite{StroudPC} in the strong disorder limit where variations in $A_{ij}$ and the flux per plaquette exert similar effects\cite{GranatoPRB86}.  For an array with $N_L$ links per plaquette and fractional variance in plaquette areas $\Delta S/\bar S$: 
\begin{equation}
\Delta A_{ij}=\frac{2\pi}{\sqrt {N_L}}\frac{\Delta S}{\bar{S}} \frac{\bar\phi}{\phi_0}
\end{equation}
The maps in Figs. 1e-h show how this disorder grows from zero with increasing magnetic field.  The phase randomization length gives the average distance that a Cooper pair travels before the gauge disorder has completely randomized its phase.  To calculate it, consider particle trajectories consisting of steps along $N$ links.  At each step, there is a random phase shift of average size $\Delta A_{ij}$, so that the distribution of the sum of the phase shifts will have a width $\sqrt{N}\Delta A_{ij}$.  When that width becomes of order $\pi$ the distribution of phases covers most of the unit circle.  This maximal phase randomization occurs on length scales of order $L_\theta=a(\pi/\Delta A_{ij})^2$ where a is the lattice constant.  Thus, $L_\theta$ provides a phase coherence length over which Cooper pair constructive interference effects can promote delocalization. The geometric disorder of the NHC array was determined by reconstructing the array with a triangulation algorithm (Figs. 1b,c).  With $\bar\phi/\phi_0=3$, $N_L=6$ and $\Delta S/\bar S=0.115$, the maximal $\Delta A_{ij}$ = 0.88 radian in our NHC films.  This corresponds to a minimal phase correlation length, $L_\theta\simeq 13a$.

Transport measurements in the low temperature limit indicate that Cooper pairs become more localized with increasing $\bar\phi/\phi_0$ (Fig. 1i).  At 130 mK, $R_{\square}$ rises monotonically by a factor of 15 for the film closest to the superconductor to insulator transition (Fig. 2a inset).  This rise spans the resistance quantum for pairs $R_Q=h/(2e)^2$ that normally separates conduction by delocalized carriers in a metallic or superconducting state from the incoherent tunneling between localized states for charge $2e$ carriers. This separation is evident in Fig. 1i as the coincident change in the temperature dependence of the resistance from $dR_{\square}/dT<0$ for $\bar\phi/\phi_0=0$ to $dR_{\square}/dT>0$ for $\bar\phi/\phi_0=1, 2, 3$.  In this Arrhenius plot, the $R_{\square}(T)$ develop an exponential dependence consistent with thermally activated tunneling with an energy barrier that increases with $\bar\phi/\phi_0$. We reproduced this evolution of $R(T)$ in a second sample on another substrate.  

We attribute this dramatic transformation in the low temperature resistance from superconducting to insulating behavior (cf. Fig. 1i) to the influence of gauge field disorder.  Ordered arrays do not exhibit this behavior.  According to experiment\cite{ZantPRL92} and the Hamiltonian in Eq. (1)\cite{kimPRB08}, a film that superconducts in zero magnetic field, superconducts at all commensurate fields. Moreover, this random gauge field tuned transition is distinct from the magnetic field tuned superconductor to insulator transitions (BSITs) that also appear in these arrays\cite{StewartPRB08,NguyenPRBRC15}.  Multiple BSITs can occur periodically at non-integer $\bar\phi/\phi_0$.  There is net vorticity at the BSITs.  This vorticity frustrates the phase ordering to make the system more susceptible to phase fluctuations.  These transitions have been observed in ordered, micro-fabricated JJAs\cite{ZantPRL92,ChenPRB95} in accord with predictions based on the Hamiltonian above\cite{fazioPR01}.  Recent work, however, has shown that the disorder in the geometry of NHC films affects the quantum critical transport at these BSITs in an unexpected manner\cite{NguyenPRBRC15}.   Here, we work at commensurate fields where there is no net vorticity but the disorder in the array geometry leads to an equal number of vortices and anti-vortices.

\begin{figure}[htb]
\begin{center}
\includegraphics[width=3.3 in,keepaspectratio]{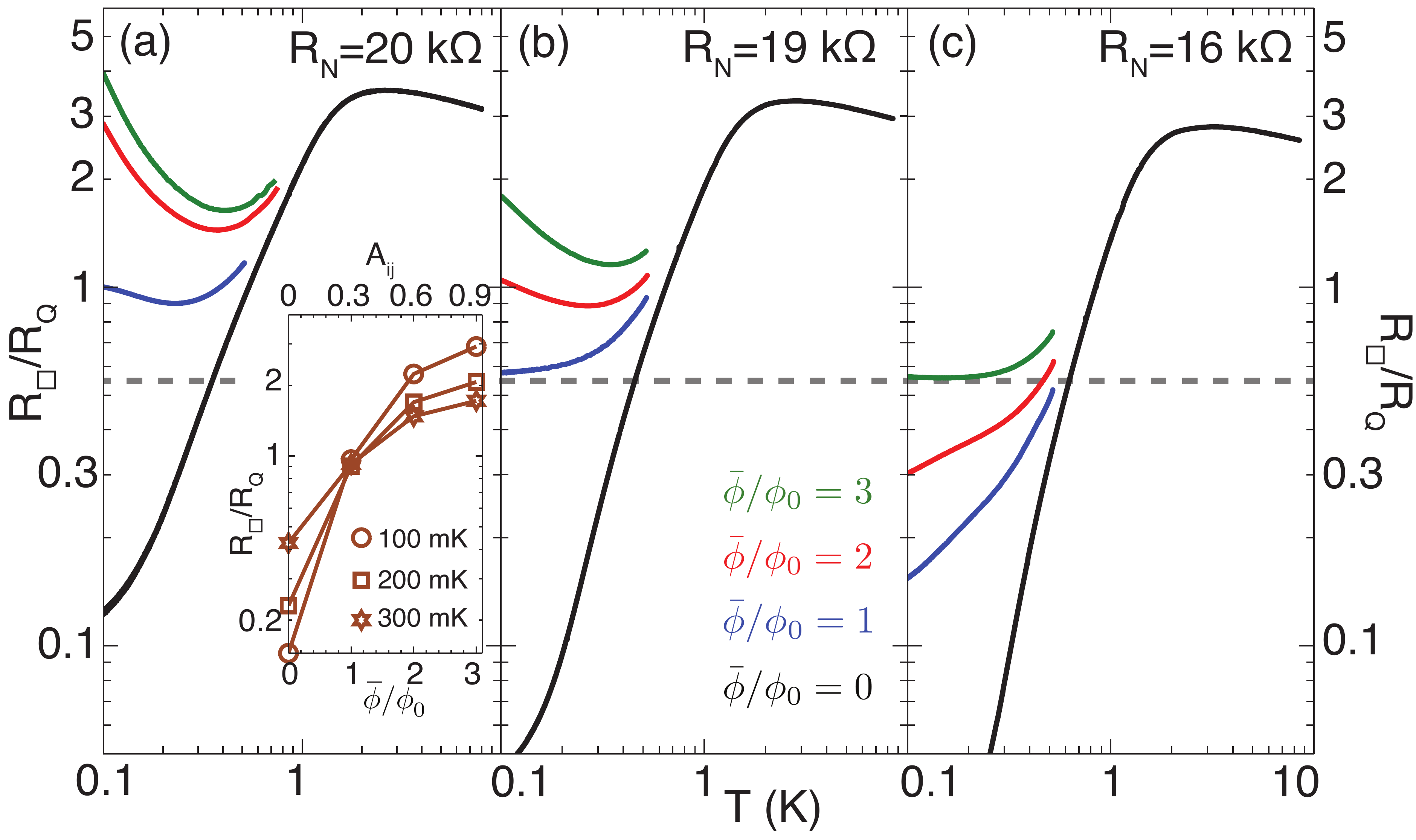}
\caption{Coupling Dependence of Random Gauge Field Tuned SIT.  a-c) $R_{\square}(T)$ of films with three different $R_N$ at 4 levels of gauge field disorder as reflected by the average flux per plaquette. The dashed line is an estimate of the critical resistance for the transition (see text). $\bar\phi/\phi_0$ = 0,1,2,3, corresponds to $A_{ij}$ = 0, 0.29, 0.59, 0.88, respectively.  Inset: $R_\square (\bar \phi/\phi_0)$ for the 20 k$\Omega$ film. }
\label{cSIT}
\end{center}
\end{figure}

The critical amount of gauge field disorder for this SIT depends on the zero field coupling constant of the film.  Figs. 2a-c shows the $R_\square (T)$ of three films with different coupling constants at commensurate fields.  They are on the same NHC substrate (Fig. 1a) so that their random gauge fields have the same magnitudes $\Delta A_{ij}$ = (0, 0.29, 0.59, 0.88) for $\bar\phi/\phi_0$ = (0, 1, 2, and 3), respectively.  To estimate the coupling constants, we presume that $J\propto T_c/R_N$ in accord with the scaling of the coupling energy of a Josephson tunnel junction and that the single island charging energy $U$ is fixed by the geometry of the substrate.  In Fig. 2, the coupling constant $K$ increases from left to right as $R_N$ decreases and $T_c$ concomitantly increases.  The $R_N$ = 20 k$\Omega$ film shows a superconducting characteristic (i.e. $dR_\square/dT>0$ as $T\rightarrow 0$) only when $\bar\phi/\phi_0=0$. It is tuned to an insulating characteristic (i.e. $dR_\square/dT<0$ as $T\rightarrow 0$) for $\bar\phi/\phi_0=1, 2, 3$.  At $\bar\phi/\phi_0=3$ the transport fits an activated temperature dependence with an activation energy of 131 mK (see Fig. 1i). The inset shows an isothermal cut of 100, 200, and 300 mK that reveals a crossing point near $\Delta A_{ij}$ = 0.3, i.e. the transition. With increasing coupling constant, the transition from superconducting to insulating behavior appears to move to higher $\bar \phi/\phi_0$ (Figs. 2b,c).  In fact, the most disordered gauge field barely tuned the  the most strongly coupled $16 k\Omega$ film into the insulating phase.  Altogether, the greater the difference between $K$ and the critical coupling $K_c$ for the zero field transition, the larger the gauge field disorder tuning parameter must be to drive the SIT.  

\begin{figure}[htb]
\begin{center}
\includegraphics[width=3in,keepaspectratio]{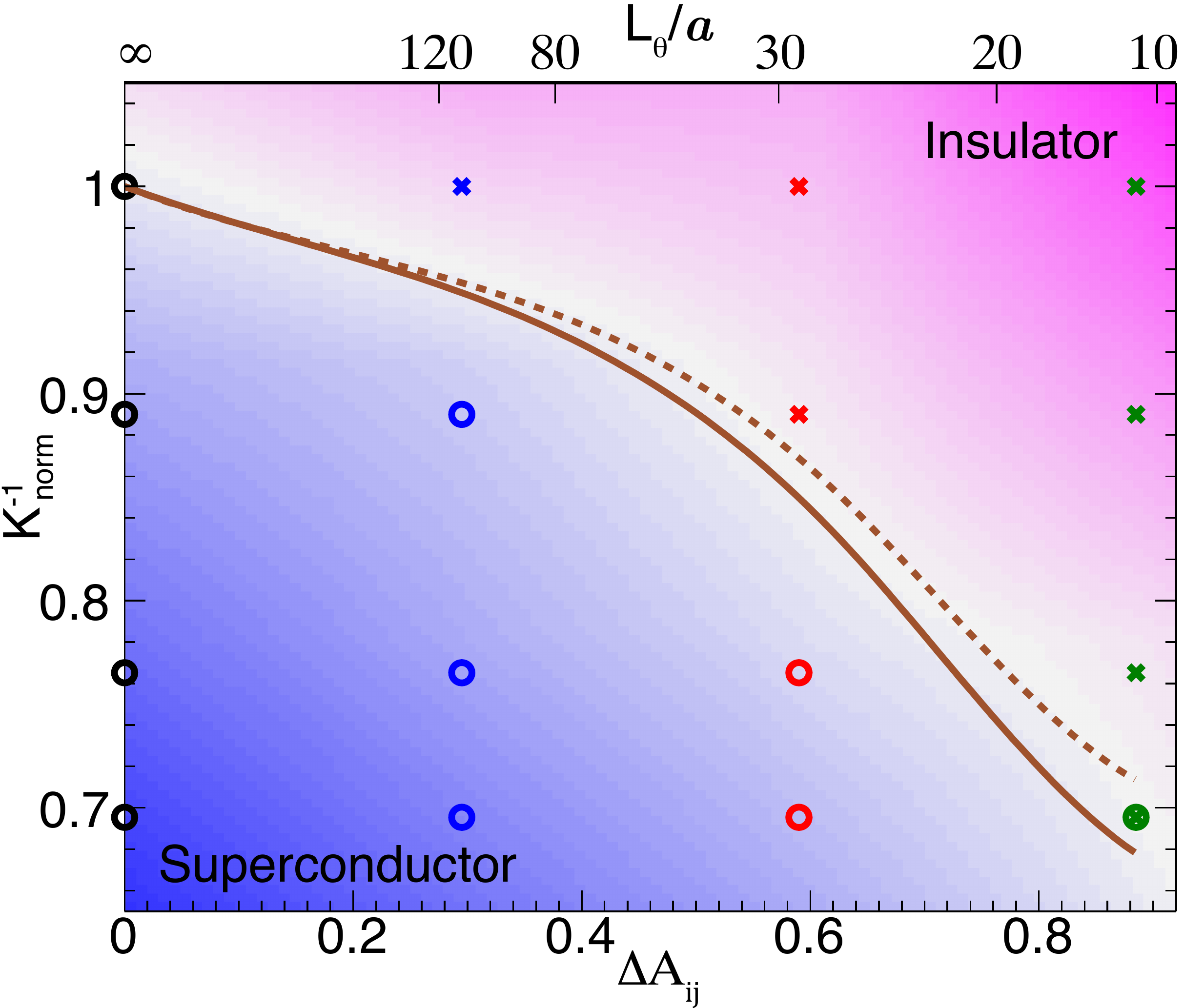}
\caption{Phase Diagram for the Random Gauge Field Tuned SIT.  Each point corresponds to a particular film with inverse coupling constant given by $R_N/T_c$ normalized to 20 $k\Omega$ film value.  Open circles correspond to films with superconducting characteristics and crosses correspond to films with insulating characteristics.  The shading is a guide to the eye for the variation of the low temperature $d log(R_\square)/dT$. The solid and dashed lines give predictions derived from figure 18 of ref.  \cite{kimPRB08} for the superconductor-insulator phase boundary without (solid) and with (dashed) magnetic pair breaking taken into account, respectively (see text).}
\label{phasediagram}
\end{center}
\end{figure}

We condense these observations into an inverse coupling constant, $K^{-1}$, versus gauge disorder, $\Delta A_{ij}$ phase diagram (Fig. 3) that enables comparison with calculations\cite{kimPRB08}.   Superconductors and insulators display positive and negative $dR/dT$ in the low temperature limit, respectively.  Any phase boundary separating them must have a negative slope indicating that the critical coupling for the transition increases with gauge field disorder.   We compare the phase diagram to Quantum Monte Carlo calculation simulations of a (2+1)D XY Model Hamiltonian (Eq. (1)) by Kim and Stroud\cite{kimPRB08}.   Those simulations showed that the ground state transforms from a phase ordered, superconducting Bose glass to a Mott insulator at a critical coupling that decreases with gauge field disorder.  The solid line in Fig. 3 correspond to their prediction matched to the data at $\bar\phi/\phi_0=0$\cite{kimPRB08}.  The dashed line corrects for magnetic field induced pair breaking effects on the coupling, which we estimated to scale as $(1-\frac{H}{H_{c2}}^2)$ appropriate for the ``dot" like structure in NHC films \cite{hollenPRB11} with an upper critical field of $\mu_0 H_{c2}= 2.5$ T\cite{Nguyen2010,nguyenPRL09}.   This comparison reveals that the data are consistent with gauge field disorder driving a Bose glass to Mott insulator quantum phase transition.   

Furthermore, as for magnetic field tuned SITs\cite{dobrosavljevic12}, there is evidence of a quantum critical resistance\cite{FisherPRL90a} at this random gauge field tuned SIT.  The low temperature tails of the $R_\square(T)$ in Fig. 2 sweep continuously from a positive to a negative slope with increasing $\bar\phi/\phi_0$ suggesting that $R_\square (T)$ becomes temperature independent at a critical amount of disorder.  A metallic flattening is most evident in the $R_{\square}(T)$ for $R_N=19$ k$\Omega$ (Fig. 2b), $\bar\phi/\phi_0=1$ and $R_N=16$ k$\Omega$ (Fig. 2c), $\bar\phi/\phi_0=3$, which appear to asymptote to 3.5 k$\Omega$. While this asymptotic separatrix is consistent with the $R_N=20$ k$\Omega$ data (Fig. 2a), none of those $R_{\square}(T)$ become level at low $T$.  We conjecture that the metallic behavior appears only at a specific coupling constant for each integer $\bar\phi/\phi_0$. The data are consistent with a critical resistance $R_c\approx 0.5 R_Q$ independent of coupling constant.  The Quantum Monte Carlo simulations that apply most directly to the present experiment predict about a factor of 3 variation in $R_c$\cite{kimPRB08} that brackets the experimental value.  The prediction that $R_c$ varies with $K$, however, is inconsistent with the data.  

In conclusion, we introduced a method to impose a random gauge field on superconducting thin films near a thickness tuned SIT.  We observed that the films can be tuned across the SIT by increasing the amplitude of the random gauge field in accord with numerical predictions\cite{kimPRB08}.  Much about this random gauge field tuned transition remains to be explored including the response of the insulator to this gauge field disorder and the discrepancy between theory and experiment on the quantum critical transport.  The capability to tune coupling and disorder independently afforded by the NHC substrate platform will be useful for such studies.  In general, experimental realizations of purely disorder driven localization transitions like the one presented here are difficult to achieve.   Theoretically, they have been studied extensively\cite{dobrosavljevic12}.  Models with potential disorder give rise to Anderson localization driven metal to insulator transitions and phase fluctuation dominated superconductor to insulator transitions, for example.  The challenge for experiments has been creating systems in which the potential disorder can be tuned independently of electron electron interactions which can also produce localization.  The development of models like that of Kim and Stroud\cite{kimPRB08} and others that employ geometrical or bond disorder\cite{dubiPRB06,swansonPRX14} and this study present new opportunities for investigations of disorder's influence on quantum phase transitions.   

\begin{acknowledgments} 
We have benefitted from discussions with D. Stroud, E. Granato, Dmitri Feldman, H. Y. Hwang, J. Joy, and Xue Zhang. We are grateful for the support of NSF Grants No. DMR-1307290 and No. DMR-0907357 and AFOSR and AOARD.
\end{acknowledgments}

\bibliography{CSIT}
\end{document}